\documentstyle[12pt]{article}
\begin{document}
\setlength{\baselineskip}{0.30in}
\newcommand{\psl}{!\!\!\!p}
\newcommand{\ksl}{!\!\!\!k}
\newcommand{\Ga}{\Gamma}
\newcommand{\la}{\lambda}
\newcommand{\be}{\begin{equation}}
\newcommand{\ee}{\end{equation}}
\newcommand{\bi}{\bibitem}
\newcommand{\al}{\alpha}
\newcommand{\ea}{\alpha_{el}}
\newcommand{\as}{\alpha_s}
\newcommand{\bb}{\beta}
\newcommand{\bef}{\beta -{\rm function}}
\newcommand{\La}{\Lambda_{QCD}}
\newcommand{\aQ}{\alpha_s(Q^2)}
\newcommand{\am}{\alpha_s(\mu^2)}
\newcommand{\aq}{\alpha_s(q^2)}
\newcommand{\ass}{\alpha_s(s)}
\newcommand{\ask}{\alpha_s(k^2_{\perp})}
\newcommand{\Gmn}{G^a_{\mu\nu}}
\newcommand{\lvac}{\langle 0|}
\newcommand{\rvac}{|0\rangle}
\newcommand{\Nc}{N_{crit}}
\newcommand{\De}{\Delta}
\newcommand{\vp}{\varphi}
\newcommand{\de}{\delta}
\newcommand{\si}{\sigma}
\newcommand{\bea}{\begin{eqnarray}}
\newcommand{\eea}{\end{eqnarray}}
\newcommand{\NP}{{\it Nucl. Phys.}~}
\newcommand{\PR}{{\it Phys. Rev.}~}
\newcommand{\PL}{{\it Phys. Lett.}~}
\newcommand{\PRL}{{\it Phys. Rev. Lett.}~}
\newcommand{\kp}{k_{\perp}}
\newcommand{\asQ}{\alpha_s(Q^2)}
\newcommand{\ls}{\lambda^2}

\begin{center}
\vglue .06in
{\Large \bf {Renormalons as a Bridge between Perturbative
and Nonperturbative Physics \footnote{Talk presented at YKIS97, Kyoto, December 1997.}.}}\\[.5in]

{\bf V.I. Zakharov}\\
[.05in]
{\it{Max-Planck-Institut fuer Physik\\
Foeringer Ring 6, 80805 Muenchen, Germany}}\\
{\it and}\\
{\it{The Randall Laboratory of Physics\\
University of Michigan\\
Ann Arbor, MI 48109-1120}}\\[.15in]

\end{center}
\begin{abstract}
\begin{quotation}

In two lectures, we overview the renormalon and renormalon-related
techniques and their phenomenological applications. We begin with a single
renormalon chain which is a well defined and systematic way to specify the
character of corrections in inverse powers of the total energy to
observables directly in Minkowski space. Renormalons demonstrate also
presence of nonperturbative contributions. We proceed then to
multirenormalon chains and argue that they are in fact not suppressed
compared to a single chain. On one hand, this phenomenon might be a
mechanism for enhancement of power corrections. On the other hand, the
derivation of relations between power corrections to various observables
becomes a formidable task and asks for introduction of models. In the
concluding, third part we consider dynamical models for nonperturbative
effects, both in infrared and ultraviolet regions, inspired by
renormalons.

\end{quotation}
\end{abstract}

\newpage

\section{Introduction }
Renormalons have various facets and the understanding of the
renormalon physics went through
various phases. The very existence of renormalons was realized within a 
pure field-theoretical framework \cite{azimov}
and by renormalons one understands usually
a simple perturbative type of graphs with a chain of many insertions
of vacuum bubbles into a bosonic line.
The renormalons were used first to discuss matters of principle
in field theory. In particular,  the renormalon chain results in 
$n!$ divergence of the expansion coefficients $a_n$ where $n$ is the
order of perturbation theory
(for a review and further references see, e.g., \cite{mueller,vz}).
Such a behaviour of $a_n$
implies in turn that the perturbative expansions are divergent and cannot
be a full answer. Thus,  nonperturbative contributions are called for.
 
Renormalon-based phenomenology emerged much later \cite{mueller,vz,sterman,
webber} and reached its peak  recently (for reviews see, e.g.,
\cite{az}). Essentially, renormalons allowed to specify
the character of power-like corrections in various observables
directly in Minkowski space. Namely,
if a physical process is characterized by a large mass scale $Q$
(like total energy in $e^+e^-$-annihilation to hadrons), then
the power corrections assume a generic form $(\Lambda_{QCD}/Q)^k$,
with k integer and depending on the variable considered.
The power corrections are encoding nonperturbative contributions.  
Renormalons are effective to find the value of $k$ in each case. 
Rather loosely, we can call it renormalon kinematics.
This kinematical part is acquiring a status of something reliable
and simple. The applications, however, are limited in nature, 
as of any kinematics.
 
A kind of renormalon dynamics shows up first once one tries to pin down 
more precisely 
what renormalons actually mean. It turns out that in fact the class
of renormalon graphs is poorly defined \cite{vainshtein}.
Namely, many renormalon chains are the same important as 
a single chain. This observation calls
for reconsideration of the results obtained in a single-renormalon
chain approximation. Generically, the power $k$ is not changed
but the relations between non-vanishing contributions 
are difficult to derive in a consistent way.
In particular, the model with the  
running coupling frozen when applied to power corrections \cite{dokshitzer}
appears to be too restrictive since it does not allow
for many renormalon chains.

Thus, at the current stage, we are invited to invent dynamical
models for non-perturbative effects 
which incorporate or are inspired by inclusion of many renormalon
chains. At the moment, it is rather a fascinating challenge than
a systematic program.  

Now, a few words on the outline of this text. The lectures at the School
were intended to be pedagogical in nature. However, 
it is rather hopeless to expose the renormalon technique in sufficient
detail within the format of 
this text. Thus, we address the reader to the reviews \cite{az}
and references therein for a systematic introduction to renormalons.
Here, we concentrate mostly on the problems encountered in
development of the renormalon physics and on the hints 
for the nonperturbtive dynamics of QCD provided by renormalons. 

\section{Single renormalon chain.}

{\bf Infrared-sensitive contributions.}

We start with a brief overview of the renormalon routine.
The general problem which the renormalons are invoked to solve
can be formulated as follows. Imagine that we consider a physical process with
large characteristic mass scale $Q$, $Q\gg \Lambda_{QCD}$,
so that the running coupling $\asQ$ is small. Then if we evaluate an
infrared safe variable, it is given as a perturbative expansion
in $\asQ$. This is the standard perturbative QCD. 
The corresponding Feynman graphs, while dominated by momenta $k\sim Q$
include also integration over
kinematical regions
which are characterized by small momenta $k\sim \Lambda_{QCD}$.
Then the effective coupling is in fact large and
there are no reliable ways to evaluate the contribution of
these, infrared-sensitive regions. It is apparently 
contaminated by nonperturbative
effects as well. The question is, how big is this
non-controllable contribution. It is rather obvious that
the price for focusing on non-characteristic momenta, $k\ll Q$, 
is a power-like suppression, $(\Lambda_{QCD}/Q)^k$.
Indeed the corresponding phase space is small. This can just be
considered a definition of what we mean by saying that
the characteristic 
momenta are of order $Q$, not $\Lambda_{QCD}$.
One can then try simple-minded estimates of the contributions coming from
the infrared-sensitive "corners" of the phase space.
Renormalons promote such estimates to a status of science. 
The power corrections, in turn, can be crucial at the pre-asymptotic
energies, or for the  precision tests of QCD.
   
{\bf Example of the thrust.}

For the sake of definiteness we will consider linear corrections
(which are leading ones) in shape variables \cite{webber,dokshitzer,az3,sterman2}. We follow Ref. \cite 
{az3} where further details of the derivation outlined below can be
found.

Consider the thrust variable which is defined as:
\be
T~=~max_{{\bf n}}{\sum_i|p_{i,{\bf n}}|\over\sum_i|p_i|}
\ee
where $p_{i,{\bf n}}$ is the component of $i$th particle momentum, $p_i$
in the reference direction ${\bf n}$ which is chosen to maximize
the right-hand side of above. Thus, for a twojet event, $T=1$ and for a 
completely symmetric 3-jet event $T=2/3$. We will consider the simplest final state
which gives rise to a non-trivial prediction for the thrust, namely,
quark-antiquark pair and a gluon, $q\bar{q}g$.
Then for soft gluons with $\omega\sim\La$ we estimate deviations from $T=1$ as
 \be
\langle 1-T \rangle_{soft}~\sim~
\int_{soft~\omega}\as{d\omega\over\omega}{\omega\over Q}~\sim~
\as{\La\over Q}
\label{lin}.\ee
Here, the proportionality to $\as\cdot(d\omega/\omega)$ 
is the standard factor due to a
soft gluon emission, while the factor $\omega/Q$ is due to definition
of the thrust since $\sum_i|p_i|\approx Q$. We cannot specify within this
simple estimate the
argument of the strong coupling constant $\as$. 
It is natural to assume, however,
that for soft gluons $\as$ is of order one.

While the very existence of the $1/Q$ corrections can be 
demonstrated in a simple way, the evaluation of the renormalon
chain, with all the coefficients fixed, is much more complicated.
It goes, however, through a well defined sequence of steps.
First, one considers one-loop perturbative calculation.

Note that the 
virtual gluon correction drops off from $(1-T)$ 
because for a two particle final state $T=1$ and we 
consider, therefore, real gluon emission, $e^+e^-\rightarrow q\bar{q}g$.
For the final state considered the thrust is known to be
\be
T~=~max(x_1,x_2,x_3)
\ee
where $x_i=2E_i/Q$ and $E_i$ are energies in the CM frame.  
For massless quarks and gluons one finds the range of the variables as:
\be
0\le x_2\le 1;~~1-x_2\le x_1\le 1;~~x_1+x_2+x_3=2.
.\ee

After a cumbersome algebra one can find that the $1/Q$ contribution to the 
thrust comes from the following term:
\be
\langle1-T\rangle_{1/Q}~=~{\as\over \pi}C_F\int_0^{1/12}
{dx\over x}\int_{T_1}^{T_2}{4\over T}
\left(1-{4x\over 1-T}\right)^{-1/2}\label{limits}
\ee
where $x\equiv\kp^2/Q^2$ and precisely this contribution gives the logarithmic
term $[ln(1-T)]/(1-T)$ in the thrust distribution.
Moreover, the $1/Q$ term is associated with small $x$,
as anticipated.
For small $x$ the limits of integration in Eq. (\ref{limits})
can be represented as an expansion:
\be
T_1~=~1-\sqrt{x}+...,~~~~T_2~=~1-4x+...~~~,
\ee
and we need to keep only the first terms in the expansion.

This calculation is still only a preliminary step to insert the
renormalon chain. The crucial point to proceed further is that
$\as$ is in fact known to run as $\as (\kp^2)$.
Therefore, it is useful to rewrite the expression for the $1/Q$ correction
as:
\be
\langle 1-T\rangle_{1/Q}~=~{2\over Q}
\int_0^{Q^2/12}
{d\kp^2\over \kp^2}\kp
\left(C_F {\ask\over \pi}\right)~
\approx~{4\over Q}\int_0^{\sim Q}d\kp\left({\ask\over \pi}C_F\right).
\label{final}\ee

{\bf Infrared renormalon in the thrust.}

Having derived expression (\ref{final}), one can 
identify the renormalon in various ways. Note first that the integral over $\kp$ in 
Eq. (\ref{final}) is obviously linearly divergent
as a function of the upper bound $Q$ 
so that, upon dividing by $Q$, the expression 
(\ref{final}) is of order $\asQ$.
But this contribution is absorbed already 
into the standard
first-order radiative corrections to the thrust. 
Now we are interested in parameterizing the IR sensitive contribution to (\ref{ld}). 

To this end we use the Borel representation for the running coupling:
\be
\ask~=~\int_0^{\infty}d\si
exp\left(-\si\left(b_0ln{k^2_{\perp}\over \La ^2}\right)\right)
\label{bc}\ee
where we assumed for simplicity one-term $\bef$,  
$b_0=(11-{2\over 3}n_f)/4\pi$.
As the next step, substitute (\ref{bc}) into (\ref{final})
and integrate first over $\kp$.
The integral over $\kp$ produces then a pole in the $\sigma$-plane
which is due to the renormalon:
\be
\left({1\over Q}\int_0^{\sim Q}\ask d\kp\right)_{renormalon}~=~
{\La\over Q}\int_{pole}{d\sigma\over -2\sigma b_0+1}
\label{ld}.\ee
To parameterize the IR contribution we are free to define the integral over pole
as, say, its principal value. In this way we come to
\be
\left({1\over Q}\int_0^{\sim Q}\ask d\kp\right)_{renormalon}~=~const {\La\over Q}\label{hadr}
,\ee
where the value of the $const$ depends on the definition of the integral over the pole and 
the procedure described can be called {\it Landau-pole parameterization}
of the power-like corrections.

Another popular way to deal with the renormalons 
is to re-express $\ask$ in terms of $\asQ$:
\be
\ask~\approx~{\asQ\over 1-b_0\asQ ln(Q^2/\kp^2)}\label{pert}~=~
\sum_n\left(b_0ln(Q^2/\kp^2)\right)^n\as^{n+1}(Q^2)
\ee
where we used the same approximation
of a single-term $\beta$-function.
The reason to re-express everything in terms of $\asQ$ is that 
eventually we are studying the perturbative expansion in $\asQ$
for the thrust. Upon substituting (\ref{pert})
into (\ref{final}) we come to the integrals of the type
\be
\int_0^Qln^n(Q^2/\kp^2)d{\kp\over Q}
~\approx~2^nn!\label{contr}
~~.\ee
Note that at large $n$ the integrand has a sharp peak at
$\kp^2\sim~e^{-2n}Q^2$.
 
Thus, renormalons produce a perturbative expansion of the type\be
\langle 1-T \rangle~\sim~\sum_{large~n}a_n(\asQ)^n~\sim~
\sum_{;arge~n}(2b_0)^nn!(\asQ )^n\label{asymp}
\ee
which is clearly divergent and 
can at best be an asymptotical expansion. 
By using the standard techniques,
it is straightforward to see that the expansion (\ref{asymp}) cannot 
approximate a physical quantity to an accuracy better than 
of order $(\Lambda_{QCD}/Q)$.
Expansion of (\ref{pert}) in $\asQ$ looks so as if we have a chain
of bubble insertions into the gluon line. Thus, the procedure can be called
{\it renormalon-chain parameterization} \footnote{The example given can actually be
somewhat misleading in the sense that the graphs with the vacuum bubble
insertions depend rather on the virtuality of the gluon, $k^2$ than
on $\kp^2$. The $\kp^2$ as the argument of the running coupling emerges
in the case considered after further algebra involving the emission
probability (see, e.g., the second paper in Ref. \cite{dokshitzer})
A more literal use of the renormalon chain can be found in the next section.} of the power-like corrections.  
Since the contribution (\ref{contr}) to $a_n$ 
comes from effective $k^2_{eff}\ll Q^2$ it is called {\it infrared (IR) 
renormalon}. 

{\bf Kinematical nature of power corrections.}

The beauty of renormalons is that, being a pure perturbative construct, they 
tell something about nonperturbative contributions as well.
Indeed, the divergence of the perturbative expansion at large $n$,
see Eq. (\ref{asymp}), implies that a unique result for the physical quantity 
can be obtained only with
account of nonperturbative contributions. In this way we learn that the
thrust receives nonperturbtaive contributions of order $\Lambda_{QCD}/Q$.
Renormalons fix in a way a minimal set of
nonperturbative terms.

On the other hand, the renormalon chain is a simple perturbative graph 
and as such is not necessarily a unique device to 
probe infrared regions. Thus, we can expect that the fixation of the value of $k$ characterizing the power
correction is a matter of kinematics and gauge invariance. Indeed, the simple 
 estimate of the infrared-sensitive
contribution to the thrust given above did reproduce the main result that
the correction is linear in $1/Q$. 

More systematically, one can utilize a finite gluon mass as a probe
of infrared sensitivity \cite{webber}. Namely, the general rule \cite{bbz}
is that terms non-analytical in the gluon mass squared, $\la^2$ are
infrared sensitive. Moreover, the correspondence between the 
non-analytical terms
and the power corrections in real QCD reads as:
\bea
a_0\as ln\ls+a_1\as{\sqrt{\ls}\over Q}
+a_2\as{\ls\over Q^2}ln\ls+a_3\as{\ls\sqrt\ls\over Q^3}+a_4\as{\la^4\over Q^4}ln\ls+...\\
\nonumber
~\rightarrow~b_0ln\La+b_1{\La\over Q}+
b_2{\La^2\over Q^2}+b_3{\La^3\over Q^3}+b_4{\La^4\over Q^4}+...\label{general}
\eea
where the coefficients $a_i$  are calculable in terms of one-loop 
Feynman graphs while the corresponding coefficients $b_i$  
incorporate the effect of large-scale fluctuations, both  perturbative
and nonperturbative, of the real QCD. 
Note that $a_0,b_0=0$ for an infrared safe quantity,
as is always assumed to be the case.

In particular, evaluation of the $1/Q$ correction to the thrust within the model with
finite $\ls$ reveals a typical source of non-analytical in 
$\ls$ terms,
that is the boundaries on the the phase space.
To find $\as \cdot(\la/Q)$ corrections to the thrust
one proceed along the same lines as above. Namely, the value of the thrust for a particular momentum configuration is 
determined by values of $x_i$, $x_i={2E_i\over Q}$, in the CM frame.
A crucial point is that  
for a nonzero gluon mass $\la$ the range of the variables 
contains $\la/Q$ terms:
\bea
0~\le ~x_2~\le~1-{\la\over Q} \\ \nonumber
1-x_2-{\la\over Q}~\le~x_1~\le~{1-x_3-\la/Q\over 1-x_2}\label{constr}
.\eea
Furthermore, one computes the thrust
distribution $(1/\sigma_0)d\sigma/dT$ by substituting the 
phase space region according to $x_{1,2,3}$ being the largest fraction and 
finding $(1/\sigma_0)d\sigma/dT$ in each region. Adding these, gives
the total thrust distribution from which $<1-T>$ is easily obtained.
The result is \cite{webber}:
\be \langle 1-T\rangle_{1/Q}~=~4C_F{\as\over \pi}{\la\over Q}.
\ee
The procedure described can be called {\it finite-gluon-mass}
parameterization of the power-like corrections.

It is worth emphasizing that
the contribution to the $1/Q$ term comes from the region 
of energetic quarks and antiquarks only and, in these regions,
from the phase space constraints (16). 
It is therefore seen to be soft-gluon dominated, 
i.e. coming from gluon with energies of order $\la$. Moreover, 
one can establish a direct correspondence between the $1/Q$ correction to the thrust evaluated via the renormalon chain (considered above) and via introduction
of finite $\la$. Namely, the identification \cite{az3}
\be
\as(\la)\la~\rightarrow~\int_0^{\sim~Q}\as(\kp^2)d\kp
\ee
brings the $\la/Q$ corrections to the form found within the renormalon technique.

This example, as well as many others, seems to indicate strongly that the 
fixation of the 
index $k$ of the power corrections in terms of renormalons is
in fact based exclusively on the consideration of phase space and of
gauge-invariance constraints. Note, however, that while the calculation with
a finite gluon mass is a straightforward one-loop calculation, the clarification
of the argument of the running coupling asks in fact for evaluation of two loops and may be a subtle matter. Imagine, for example, that
the argument of $\as$ would be the invariant mass of the quark-gluon system
$\as(m^2_{qg})$ rather than $\ask$. Then for soft gluons $m^2_{qg}\sim\Lambda^2_{QCD}$
only for $\omega_g\sim\Lambda^2_{QCD}/Q\ll \Lambda_{QCD}$, not
$\omega_g\sim\Lambda_{QCD}$ as discussed above. Then, there would be no linear corrections at all.

{\bf Open questions.}

There exist at the moment open problems which arise already on this, kinematical level of the renormalon calculus. In a way, they are reflections of the same question, whether different infrared probes give the same answer. We will
spell out in some detail two open questions:

(a) In a single renormalon approximation, power corrections can be studied
in case of deep inelastic scattering as well \cite{stein}. In view of the richness of the
data, these predictions may provide an important test of the theory
(for a review and further references see the third paper in Ref. \cite{az}).
However, there exists an unresolved controversy over the predictions themselves \cite{az2}.
In more detail, one considers, as usual, the amplitude for $\gamma^{*}+q\rightarrow q^{'}+g$
where $\gamma^{*}$ is the virtual photon with 4-momentum $q$  and q is the parton of momentum $p$ but keeps the gluon mass $\la$ finite.
Then the leading non-analytical in $\ls$ contribution to, say,
the longitudinal structure function $F_L$ takes the form \cite{az2}:
\be 
(F_L)_{\ls}~=~C_F{\as\over 2\pi}4z^2{2\ls ln\ls\over Q^2}\label{stein}
,\ee
where $z~=~Q^2/2(p\cdot q), ~q^2=-Q^2$. If one replaces the $\ls ln\ls$ term
by $(\Lambda^2_{QCD}/Q^2)$
according to the rules discussed in the preceding subsection,
then the result coincides with the evaluation of the renormalon chain
performed earlier \cite{stein}. The most interesting point about
(\ref{stein}) is that the $z$ dependence is fixed and this $z$ dependence
of the power correction can be checked experimentally.

On the other hand, if one uses the observation that the coupling runs 
as $\ask$ and applies the same technique as above then the resulting
$z$-dependence is different:
\be
(F_L)_{1/Q^2}~=~{C_F\over 2\pi}{4z^2\over 1-z}{\Lambda^2_{QCD}\over Q^2}I
\ee
where the explicit expression for the integral $I$ can be found
in the original paper \cite{az2}. 

While the both derivations are straightforward, no understanding of this discrepancy was reached so far.

(b) In some cases the leading $1/Q$ corrections cancel. 
The best known and first discovered \cite{bb}
example of this kind is the Drell-Yan (DY) process. 
The difference between, say, thrust and the DY cross section
can be traced \cite{az4} to the difference in resolutions
needed to measure the corresponding observables. Namely, in case
of the thrust the momenta of individual particles are assumed
to be measured with the precision of order $\Lambda_{QCD}$
while in case of the DY process the resolution does not involve
$\Lambda_{QCD}$. The proof of this assertion \cite{az4} utilizes the
Kinoshita-Lee-Nauenberg and Low theorems. Now, an unsolved problem \cite{asz} is that inclusion
of higher loops brings mixture of soft and collinear divergences
.
As a result the Low theorem holds naively only for frequencies
$\omega< m_q^2/E$ where $m_q$ is the quark mass and $E$ is its energy.
In the abelian case it can be extended to $\omega\sim m_q$
\cite{dd}. In the nonabelian case such an extension has not been proven so 
far. Thus, in case of the Drell-Yan cross section there is no proof that
the index $k$ does not change from $k=2$ to $k=1$ once two-loop corrections are included. 

{\bf Summary to part I.}  

We can summarize the presentation above by saying that there exist well defined and systematic techniques to parameterize infrared sensitive contributions to the infrared safe quantities in terms of the power 
corrections $(\La/Q)^k$. Despite some open questions the techniques might well find their way to textbooks soon.
In the next section we will argue, however, that the renormalons are not exhausted
by a single chain.

\section{Multi-Renormalon Chains.}

In perturbative QCD  observables are
calculable as a series in $\asQ$, $f=\sum_na_n(\asQ)^n$. As far as $\asQ$ is small enough, each next term in this expansion is small compared to the previous
one (we should qualify this statement, in fact, for $n$ to be not too large).
The power-like corrections revealed by renormalons can be viewed as
representing kinematical regions near the boundaries of the phase space.
One might think about them as of a small part of the corresponding perturbative
contribution. However, the basic observation on the multi-renormalons 
is that the power-like corrections
associated with higher loops of the perturbation theory are in
fact not suppressed at all. One may say that the whole perturbative expansion collapses once projected onto the power-like corrections. This phenomenon was observed first on the example of ultraviolet renormalon \cite{vainshtein} and reiterates itself in 
other cases as well \cite{faleev,az5,az6} (for related discussions see also
\cite{also}). In the subsequent subsections we will consider examples of the heavy-quark potential at short distances \cite{az6}, ultraviolet renormalons \cite{vainshtein}, dispersive approach to the running coupling \cite{az5}. All these examples are rel
evant also to discussion of dynamical models in the next section.

{\bf A simple estimate.}

The origin of the phenomenon of the "collapse" of the renormalon chains can be understood in terms of the  simple estimate of the soft particle contributions to the thrust given at the
beginning of the previous section. Namely, let us consider now 
two soft gluons instead of a single gluon. Neglecting the
nonabelian nature of the gluons, we may then write:
\be
\langle 1-T\rangle_{soft}~\sim
\int_{soft~\omega_1}\as{d\omega_1\over \omega_1}\int_{soft~\omega_2}
\as{d\omega_2\over \omega_2}{\omega_1+\omega_2\over Q}~\sim~{\Lambda_{QCD}\over Q}
,\ee 
i.e., the correction is of the same order as for a single soft gluon (see Eq. (\ref{lin})). Two effects are combined to escape possible suppressions. First, the effective coupling constant is of order unit since we are dealing with soft emission. Second, 
the origin of $\Lambda_{QCD}/Q$ correction is pure kinematical. Namely, the deviations from $T=1$ because of the emission of a soft particle is of order $\omega/Q$. 
This is the price for considering an unnatural-scale momenta, $\omega\sim\Lambda_{QCD}$
(and not $\omega\sim Q$). It is crucial therefore that this price is paid only once, independent of the number of soft particles. 

{\bf Heavy-quark potential at short distances.} 

A weak point of the estimate in the preceding subsection was the assumption
that the coupling is of order unit. One can circumvent this difficulty,
as we will argue here on the example of the heavy-quark potential at short distances \cite{az6}.

As the distance $r$ between the quark and antiquark tends to zero,
$r\ll\Lambda_{QCD}^{-1}$, the
potential becomes Coulomb-like:
\be
\lim_{r\rightarrow~ 0}{V(r)}~=~-{C_F\over r}\sum_na_n\as^n(r)+
\sum_nv_n\La^{n+1}r^n.\label{potent}
\ee
where $a_n,v_n$ are expansion coefficients ($a_1=1$) and we reserved for power-like corrections as well.

To pick up the renormalon contribution, it is convenient to start from the
momentum representation. To first order in the running coupling:
\be
V(r)~=~-C_F\int{ d^3{\bf k}\over (2\pi)^3}(4\pi\as({ k}^2)){exp(i{\bf k\cdot r})\over {k}^2}\ee
where (only in this subsection) $k^2\equiv {\bf k}^2$. As the next step we are using again the Borel representation,
$\as ({ k}^2)~=~\int d\sigma\left({k\over\La}\right)^{-2\sigma b_0},
$
to get
\be
V(r)~=~-{2C_F\over \pi}\int_0^{\infty}d\sigma \int_0^{\infty}
dk{sin (kr)\over kr}\left({k\over \La}\right)^{-2\sigma b_0}.\label{irpot}\ee
 Moreover, since we are interested in small-$r$ behaviour we expand the sine in $kr$
and get \cite{agl} for the corrections to the Coulomb-like potential:
\be
\delta V(r)~=~v_0\La+v_2\La^3r^2+...~~~~~\label{laqcd}
.\ee
It is noteworthy that odd powers of $r$ are absent for this single renormalon chain. We shall come back to this point in the next section.

Our central point now, however, is the contribution of many renormalon chains.
This can be extracted in a remarkably simple way once perturbative
expressions for $V({k}^2) $ are available in higher loop approximation.
In fact first three terms in the purturbative expansion were calculated explicitly \cite{peter}:
\be
V(k^2)~=~-{4C_F\pi\alpha_{\overline{MS}}\over {{ k}}^2}
\left(1+{\alpha_{\overline{MS}}\over \pi}(2.583-0.278n_f)+
{\alpha^2_{\overline{MS}}\over \pi^2}(39.650-4.147n_f+0.077n_f^2)\right)
\ee
Moreover, in the approximation of the one-loop $\beta$-function,
\be\as^2({{k}}^2)~=~{1\over 2b_0}\La{d\over  d\La}\as({k}^2).
\label{reduction}\ee
and by differentiating $n$ times Eq. (\ref{laqcd}) with respect to $\La$ 
we immediately find contribution of $(n+1)$ renormalon chains, 
associated with the $\as^{n+1}({{k}}^2)$ in the perturbative expansion.

Thus, we have a unique possibility to compare the 
contributions of one-, two- and three-renormalon chains:
\be\de V(r)~\approx~(const)\La^3r^2(1+1.1+6.0+..).\label{numbers}\ee
. However, there is no convergence of the series in sight in that case either.

To summarize, we have shown that the power corrections associated with many
(infrared) renormalon chains are not suppressed compared to the power
corrections introduced via a single renormalon chain. 

{\bf Coupling freezing vs renormalons.}

The picture of infinitely many renormalon chains, 
all of which are equally important, is difficult to be transformed to a 
well defined model. 
The simplest assumption is that actually a single chain reproduces correctly the nonperturbative dynamics of QCD. There are well known models to this effect.  We have in mind, first of all, the model with freezing of the running coupling in infrared at a 
low value.
If the effective coupling is always small, then two soft particle emission (see the discussion
at the beginning of the section) is suppressed  and the major uncertainty in theoretical predictions disappears. One may relate then various $1/Q$ corrections in much more straightforward way
\footnote{In fact,
as is argued in Ref. \cite{twoloop}, two-loop effects are not suppressed within this model, while higher loops are negligible. For simplicity of presentation, we do not discuss this complication.}.
In abelian case, one can use the model with a finite gluon mass
(see section 2) in higher-loop approximation. Since the emission of soft
abelian gluons is independent this model amounts again to the dominance of a single renormalon chain.

Thus, one may say that these models contradict our relations between, say,
one and two renormalon chains.
It is easy to understand the reason
for this discrepancy. Namely, the renormalon technique utilizes the fact that there is only
one dimensional parameter in QCD, that is $\La$. Models with dominance of a single
chain introduce explicitly or tacitly other mass scales. In the cases mentioned above it is either the gluon mass or the scale where the coupling freezes out.

The phenomenology of the power corrections is 
rich enough to distinguish between various models. In particular, let us concentrate on the thrust $T$ (see section 2) and heavy and light jet masses, $M_h^2$ and$M_l^2$, respectively. Then, the
one-chain dominance implies \cite{dokshitzer,az3}:
\bea
<1-T>_{1/Q}~\approx~\langle{M^2_h\over Q^2}\rangle_{1/Q},\\ \nonumber
\langle{M_l^2\over Q^2}\rangle_{1/Q}
~\ll~\langle{M^2_h\over Q^2}\rangle_{1/Q}.  
\eea
Indeed, in the one loop approximation, one jet becomes "heavy",
$M_h^2\sim\asQ Q^2$ while the other jet remains "light", i.e. represented by a bare quark. Moreover, the pattern of the power-like corrections repeats the pattern of one-loop perturbative terms since it is a part of this correction. The latter statement i
s especially obvious in case of the non-zero gluon mass.

To summarize, the model with freezing of the running coupling, as applied to the power corrections, is in variance with the conclusion on importance of the multi-renomalon chains. Moreover, it leads to a well defined pattern of power corrections and can b
e tested experimentally. We shall discuss this point further in Section 3.

{\bf Ultraviolet renormalon.}
 
So far we considered contributions of momenta much smaller than the typical
mass scale $Q$. Integration over virtual momenta brings contributions
of momenta much larger than $Q$ as well. 
These are detected through ultraviolet renormalons
which are discussed in more detail in this subsection.

Imagine that we are considering a Feynman graph which is convergent and
saturated by typical momenta of order $Q$. If we are interested 
specifically in
the contribution $\de_{UV}$ of virtual momenta $k^2> Q^2$,
then we could start with the following estimate:
\be
\de_{UV}~\sim~Q^2\int_{\sim~Q}^{\infty} {d^4k\over k^6}\as(k^2)\label{naive} 
\ee
where we put the factor $Q^2$ to make the quantity dimensionless.
For example, we may always consider the fraction of the total answer which is due to very large momenta. 
Note, however.
that because because the coupling depends on $k^2$ only logarithmically, the integral (\ref{naive}) is saturated at its lower bound by $k^2_{eff}\sim Q^2$ and
we would end up with consideration of the same typical momenta of order $Q$.

The situation is changed if we substitute the expression for $\as(k^2)$
as an expansion in $\asQ$ and consider large-order terms in this expansion.
Then for the coefficient $a_n$ of the perturbative expansion in $\asQ$ we
get the integral:
\be
a_{n+1}~\sim~b_0^n(-1)^nQ^2\int_{\sim~Q}^{\infty}{d^4k\over k^6}
\left(ln{k^2\over Q^2}\right)^n~\sim~(-1)^nb_0^n\cdot n!\label{uv}
.\ee
which is dominated by $k^2_{eff}\sim e^n Q^2$, where
$e$ is the base of the natural logs. Thus in this case $k^2_{eff}\gg Q^2$
and we encounter the {\it ultraviolet renormalon}. 

To get a rough estimate of the effect of two renormalon chains we consider 
an expression similar to
(\ref{naive}) but with $\as(k^2)$ replaced by $\as^2(k^2)$. Applying again 
Eq. (\ref{reduction}) we can readily see that two renormalon chains
result in the same asymptotical behaviour of the coefficients $a_n$ at
large $n$ as a single chain. 

Such calculations can be performed in much greater detail \cite{vainshtein}.
In particular two-loop effect can be accounted for systematically,
also in the $\beta$-function. The results are in agreement with the estimates above.
Moreover, one can show that the "towers" of momenta, $k^2_2\gg k_1^2\gg Q^2$
still result, generally speaking, in the same asymptotical behaviour of $a_n$ as above.
Instead of going into technical  details, let us mention another simple
way to understand the effect. Namely, one might argue that the renormalon
chain is singled out because, to a given order $n$, the power of
the log in the integrand (see Eq. (\ref{uv})) is the highest possible. 
Moreover, since logs are eventually large, $ln(k^2_{eff}/Q^2)\sim n$ this contribution
appears to be the largest. Indeed, two renormalon chains produce one power of the log less. However, the two-chain contribution is enhanced combinatorially.
The reason is that now the same number of the bubble insertions, $n$, can be distributed between two chains. As is readily seen, this combinatorial factor is also of order $n$. Thus, two mechanisms of the growth of $a_n$ with $n$, namely large combinatori
al factors and large logs in fact get mixed up. 

Note that so far we have been considering the perturbative expansion in
$\asQ$ which is most natural.
However, one might consider expansion in the coupling normalized at arbitrary point $\mu^2$ as well. For example, expansion in the bare coupling 
$\as(\Lambda^2_{UV})$ is relevant to the lattice studies (with $\Lambda_{UV}\sim a^{-1}$)
\cite{pino}.
Then the perturbative series is reshuffled considerably. In particular, in terms
of the expansion  in $\as (\mu^2)$ the ultraviolet renormalons are power-like
suppressed if $\mu^2\gg Q^2$ \cite{bz} and disappear completely if
$\mu^2=\Lambda^2_{UV}$. Indeed, there is
no integration over virtual momenta $k^2> \Lambda^2_{UV}$ at all.
Instead of (\ref{uv}) we now have:
\be
Q^2\int_Q^{\Lambda_{UV}}\as(k^2){d^4k\over k^6}~\approx~{Q^2\over \Lambda_{UV}^2}\sum_{n\sim 1}^{\sim N_Q}(b_0)^nn!\as^n(\Lambda^2_{UV})~~~, N_Q~=~ln(\Lambda_{UV}^2/Q^2).
\ee
Thus, we have a power suppressed, as $Q^2/\Lambda^2_{UV}$, term with a series in $\as(\Lambda^2_{UV})$ in front which behaves as an infrared renormalon in an unusual
position and with $n!$ behaviour tempered at $n\approx N_Q$. Thus, in terms of the expansion in $\as (\Lambda^2_{UV})$ the ultraviolet renormalon appears to be related to the power suppressed terms of order $Q^2/\Lambda^2_{UV}$. There are first indication
s on the existence
of terms $\sim( \La\cdot a)^2$ in the lattice simulations \cite{pino} but the
correspondence (if any) with UV renormalons has not been considered in any detail. 

{\bf Effective four-fermion interaction.}

The UV renormalon is most effectively treated within an operator
product expansion which utilizes the observation $k_{eff}^2\gg Q^2$
\cite{parisi,vainshtein,kivel}. Moreover, since
the ultraviolet renormalon is related to very high momenta, the structure
of this OPE is universal. The leading operators are
singled out through their anomalous dimensions and various channel differ
only in values of matrix elements of these operators  \cite{kivel,vainshtein}.

It is remarkable that at the level of two chains there appear four-fermionic
operators which actually dominate the large $n$ behaviour of the expansion
coefficients $a_n$ \cite{vainshtein,peris,kivel}.
For purpose of orientation, one may keep in mind that the dominant
operator $O_6$ is of the form:
\be
c(k^2)O_6~=~{\sum_n\as^n(k^2)\over k^2}\left(\bar{q}t^bq\bar{q}t^bq+\bar{q}t^b
\gamma_5q\bar{q}t^b\gamma_5q\right).
\ee
where $t^b$ is the Gell-Mann matrices in flavor space and $k^2$ is the effective $k^2_{eff}$ associated with ultraviolet
renormalons, $k^2_{eff}\gg Q^2$ (see above). Moreover, $c(k^2)$ is a coefficient
function which can in principle be obtained as an expansion in $\as(k^2)$.
Although there are many similarities of the technique with the standard
operator product expansion, there are important differences as well.
In particular, the matrix elements of the operators can be calculated perturbatively as far as $Q$ is large while truncation of the expansion for $c(k^2)$
at any finite order in $\as(k^2)$ changes the resulting coefficients $a_n$ by order unit (for details see \cite{vainshtein}). 

It is remarkable that similar four-fermion effective interactions
are introduced in Nambu-Jona-Lasinio model, with $k^2$ replaced by a mass
scale of order $\La^2$. It is for the first time that a four-fermion
effective interaction is generated by well defined graphs in fundamental QCD and one may speculate further on matching of the UV renormalon with the NJL model at an intermediate mass scale \cite{yamawaki}.

{\bf Dispersive approach to the running coupling: $1/Q^2$ terms.}

The basic feature of the ultraviolet renormalon is that the
corresponding perturbative
series is Borel summable.
Indeed, a simple function of the running coupling:
\be
f(\asQ)~=~\int_0^{\infty}dt{exp(-{t/\asQ})\over (1+b_0t)}\label{borel}
\ee
has the same perturbative expansion as produced by the ultraviolet
renormalon. The function (\ref{borel}) is the Borel sum of this expansion.
Note that the sign alternation of the expansion coefficients, 
$a_n\sim(-1)^n$ is crucial for the Borel
summability.

If $a_n\sim(-b_0)^nn!$
then the absolute values of the subsequent terms in the perturbative sum fall down until
a critical value $N_{cr}$ is reached:
\be
N_{cr}~\approx~{1\over b_0\asQ}
.\ee
For $n>N_{cr}$ the values of $|a_n(\asQ)^n|$ grow with $n$.  
The Borel summation replaces this, growing branch of the perturbative
expansion by terms of the order $a_{N_{cr}}(\asQ)^{N_{cr}}$. In other words, the Borel summation effectively introduces a nonperturbative term of order $1/Q^2$:
\be\de_{non-pert}~\sim~{\La^2\over Q^2}
.\ee

Although such a term can be small numerically and subordinate to many terms
in the perturbative expansion with $n<N_{cr}$, it brings in a problem of interpretation \cite{vz}. Indeed, in the spirit of the quark-hadron duality the $1/Q^2$ correction
would be naturally associated with contribution of low-lying hadronic states
since the corresponding imaginary part is proportional to $\de (Q^2=0)$ and should be smeared over $Q^2\sim\La^2$ to imitate the
hadronic contribution. On the other hand, as we saw in the preceding subsections the ultraviolet renormalon is associated with huge virtual momenta, $k^2_{eff}\gg Q^2$. Thus, the duality looks unnatural. Note that in case of the infrared renormalons the s
ituation is very different and natural: IR renormalons are dual to
low-lying hadronic states. Similar basic idea underlies the QCD sum rules
\cite{svz} (for a review at this Workshop see Ref. \cite{shifman}). The ultraviolet renormalon defies this logic by introducing duality between low-lying hadronic states and very large virtual momenta in perturbative graphs.

This kind of connection between the imaginary part at low $Q^2$ and real part at asymptotically large $Q^2$, arises if one postulates dispersion relations.
In the context of the ultraviolet renormalon, the dispersion representation
is to be used for the running coupling itself. Then $1/Q^2$ terms are introduced in the definition of coupling \cite{grunberg,az5} (analytical properties of the running coupling can be of course discussed without any connection to renormalons, see, e.g., 
Refs. \cite{aswell,dokshitzer}).

The connection between IR and UV momenta is realized then in the following way. In perturbation theory, the coupling satisfies dispersion relations and
the corresponding cuts are at $s>0$. If we assume analyticity for a resummed coupling as well then the standard expression, $\asQ\approx 1/b_0ln(Q^2/\La^2)$, contains a pole at $Q^2=\La^2$. On the other hand, no finite order of perturbation theory has a c
ut below $s=0$. The non-vanishing imaginary part at the
tachyonic momenta is thus introduced via the summation procedure of all the orders in $\asQ$. The
summation procedure is well known to be justified in the leading log approximation and is perfectly consistent at this level. 

It is an open question how to define the coupling to the accuracy
$\La/Q^2$. In particular, one may introduce a running coupling with the Landau ghost removed \cite{aswell}:
\be
\overline{\asQ}~=~{1\over b_0ln(Q^2/\La^2)}+{\La^2\over b_0(\La^2-Q^2)}.
\label{effcoup}\ee
At large $Q^2$ the coupling $\overline{\asQ}$ differs from $\asQ$ by a negative $1/Q^2$
correction. Note that it is the tacit assumption on the analyticity that 
relates the removal of the Landau pole to the $1/Q^2$ correction at large $Q^2$.

The crucial point which we are emphasizing again and again in this section
is that higher loops in perturbative expansion collapse to the same order
of magnitude power corrections. Within the current set up of the analyticity,
consider the effect of removing the Landau pole from $\as^n(Q^2)$.
Note that we should remove the single pole:
\be
{1\over b_0^n ln^n(Q^2/\La^2)}~\rightarrow~
{1\over b_0^n ln^n(Q^2/\La^2)}-{\La^2\over (n-1)!b_0^n( Q^2-\La^2)}.\label{removal}
\ee
and we come again to a $1/Q^2$ correction which is not suppressed
by any power of $\asQ$. In particular, if we have an ultraviolet renormalon then
\be
\sum_nn!(-b_0
)^n\as^{n+1}(Q^2)~\rightarrow~-{\La^2\over b_0 Q^2}\sum_n(-1)^n
\ee
as far as the $1/Q^2$ corrections are concerned.
To make sense out of the latter sum we can introduce again the Borel summation  
(which sums it up to $1/2$). In this way we see again that the Borel summation
of the UV renormalon and introduction of the $\La^2/Q^2$ corrections to the
coupling are interrelated. A condition for the consistency of the two approaches might well exist but has not been investigated.

Eq. (\ref{removal}) demonstrates \cite{az5} that the procedure of the removal
of the Landau pole does {\it not} reduce to a universal redefinition of the coupling on the $1/Q^2$ level (as is usually tacitly assumed). Instead, the $1/Q^2$ correction depends on all the terms in perturbative expansions in a nontrivial way.

{\bf Summary to part II.}

To summarize, we have demonstrated that the power suppressed terms receive contributions of the same order from higher loop graphs. The observation can be phrased different way as well. Namely, in case of the logarithmic terms (pure perturbative QCD) larg
e logs are known to cancel between virtual and real gluons as far as inclusive observables are concerned. Because of these cancellations higher loops are suppressed by extra powers of a small coupling $\asQ$. No such cancellations are known for the power-
suppressed corrections. 

The accumulation of the contributions due to higher loops could provide a 
pure perturbative mechanism for an enhancement of the power corrections.
However, it is not known at all whether this enhancement is actually taking place.
And even if one is prepared to speculate that the enhancement is there, experimental consequences of this hypothesis are not clear since there is no model independent way to relate various channel. We will consider some models in the next section.

\section{ Dynamical models.}

{\bf Infrared renormalons and hadronization models.}

It seems rather obvious that the IR renormalons should converge
with simple hadronization models. Indeed, if we look afresh at, say,
Eq. (\ref{hadr}) then we immediately recognize renormalons as introduction
of "intrinsic" transverse momentum for partons, or the tube model
(for a review see Ref. \cite{webber3}). Moreover, the same $1/Q$ corrections were first discussed within this kind of an approach long ago \cite{barreiro}.
However, it is not straightforward at all to quantify this feeling and
fix the relation between renormalons and hadronization models in any rigorous way. The problem is the same as in the previous section: on one hand, all higher loops become equally important for the power corrections and, on the other hand, there is no han
dle to treat
all the terms in perturbative expansions.  

The statement on the universality of the $1/Q$ 
 corrections \cite{az3,sterman2} falls closest to fixation of the relation between the renormalons and hadronization models. To derive the universality one ensures first that nearly twojet configurations dominate an observable. This can be achieved in a p
ure kinematical way. For example, in case
of the thrust one introduces the following average \cite{sterman2}:
\be
\langle exp(-\nu(1-T))\rangle_{1/Q}~=~exp(-2\nu U/Q)\label{defin}
\ee
where $\nu$ is a large number and $U$ parameterizes infrared region.
The dynamical question is how to find $U$.

In the universality framework and to first order in $\ask$:
\be
U~=~{C_F\over\pi Q}\int_0^{\sim ~Q}{d\kp^2\over \kp}\ask~\sim~{\La\over Q}
\label{u}.
\ee
Although Eq. (\ref{u}) looks very similar to the single chain approximation
they are not identical. Namely, there is a crucial factor of 2 in the value of $U$. The justification for this factor of 2 is, roughly speaking,
as follows. We start with perturbative graphs and pick up the contributions which contain logs and dominate the thrust distribution
(see the remark following Eq. (\ref{limits})). Then we continue these contributions to the infrared sensitive region using the renormalon technique. The factor of 2 is due to the fact that originally there are large
contributions associated with gluons aligned with each quark.
Note that the momentum configuration which finally determines the $1/Q$ correction corresponds to a soft gluon, not a collinear one. Hence, there
is no solid proof of this factor of 2 \cite{nason}. 

With inclusion of higher loops, the universal factor $U$ becomes
\cite{az3,sterman2}:
\be
U~=~\int_0^{\sim~Q}{d\kp^2\over Q\kp}\gamma_{eik}(\ask),
\ee
where $\gamma_{eik}$ is the so called cusp anomalous dimension. This
substitution allows for extrapolation to the infrared-sensitive region of all the log
terms in the perturbative calculations. Since these logs are known to be universal the power corrections turn universal as well.
The universality of the $1/Q$ corrections implies,
in particular, that the same quantity $U$ enters the expressions for other shape variables, e.g. \cite{az3}:
\be
{1\over 2}\langle 1-T\rangle_{1/Q}~=~{1\over 3\pi}\langle C\rangle_{1/Q}~=~
{2\over \pi}\langle{\sigma_L\over \sigma_T}\rangle
_{1/Q}~=~{1\over \pi}\langle Esin^2\delta\rangle_{1/Q}~=~U.
\ee
Here we used the standard notation for various shape variables
and for simplicity did not introduce the exponential weighting
(see Eq. (\ref{defin})).

From the phenomenological point of view the predictions from the universality
of the $1/Q$ corrections are close to the tube model. In particular, we expect \cite{az3}:
\bea
\langle 1-T\rangle_{1/Q}~=~\langle{M_h^2\over Q^2}\rangle_{1/Q}
+\langle {M_l^2\over Q^2}\rangle_{1/Q}\\ \nonumber
\langle {M_h^2\over Q^2}\rangle_{1/Q}~\approx~\langle{M_l^2\over Q^2}\rangle_{1/Q}
\eea
which is to be contrasted with an alternative set of predictions 
(28) based on the assumption on the dominance of a single renormalon chain.
It is our understanding that the data \cite{delphi} clearly favor 
the multi-renormalon, or universality models but the fits seem to be subject to large
systematic uncertainties.

{\bf UV renormalons and short distance physics.}

While everybody is prepared to consider models for infrared phenomena,
treatment of the ultraviolet region is expected to be rigorous. Indeed,
the effective coupling is small and the nonperturbative effects  
arguably vanish as a high power of $Q^{-1}$ \cite{nsvz}.
Ultraviolet renormalons are associated, on the other hand, with
$\La^2/Q^2$ corrections. Thus,  
if we assume that UV renormalons give indeed a significant contribution
then we introduce in fact new nonperturbative effects at short distances.
Let us first substantiate this simple point.

Consider to this end again the heavy quark potential at short distances.
If we allow for a $1/Q^2$ piece in the effective coupling 
(see Eq. (\ref{effcoup}) above) then we have for deviations of the
potential from the Coulombic one at short distances:
\be
\lim_{r\rightarrow ~0}{\de V(r)}~\sim~\int d^3{\bf k}
exp(i{{\bf k\cdot r}}){\La^2\over {\bf k}^4}~\sim ~\La^2\cdot r
.\ee
Such a linear piece does not arise from the infrared region at all.
Moreover, the heavy quark potential is measurable on the lattice 
(see, e.g., \cite{bali}) and
in this sense is observable. Let us emphasize that we are discussing short distances while the linear potential is of course very common at large distances but is expected to change to a $r^2$ correction at short distances
(see Eq.(\ref{irpot})) according to the conventional picture (see discussion
in Section 3). So far, this prediction has not been checked.
Moreover, there are first indications of the $1/Q^2$ correction to the
effective coupling obtained on the lattice \cite{pino}. Thus, it might be right time
to look for implications of the $1/Q^2$ corrections associated with short distances.

To visualize what kind of new physics we can deal with let us
consider a simple estimate of nonperturbative effects in $V(r)$ at short distances \cite{az6}. Namely, let us represent the potential energy of the $Q\bar{Q}$ pairin an abelian case as an integral over space from the quark electric field:
\be
V(r)~\sim~{1\over 4\pi}\int d^3{\bf r}^{'}{\bf E}_1({\bf r}^{'})\cdot{\bf E}_2({\bf r+r}^{'})\label{perfect}
.\ee
As far as we substitute the electric fields ${\bf E}_{1,2}$ as the fields of point-like charges, the Coulomb potential is perfectly consistent with (\ref{perfect}) of course. Imagine, however, that the electric field is changed at large distances because 
of the confining properties of the vacuum. Then there would
be a corresponding change in the potential at short distances as well.
Moreover, let us assume that strong perturbative fields are allowed in the vacuum while weak fields below some critical value are not allowed:
\be
{\bf E}^2_{cr}~\sim~\La^4 ~~
or~~
{\as r^2\over R_{cr}^6}~\le\La^4
\ee
where we accounted for the fact that at the distances large compared to $r$
the field of the $Q\bar{Q}$ pair is that of a dipole.
As the last step, estimate the change in the potential at short distances due to the screening of the field at $r>R_{cr}$:
\be
\lim_{r\rightarrow~0}{\delta V(r)}~\sim~\as r^2\int_{R_{cr}}^{\infty}{d^3r^{'}\over (r^{'})^6}~\sim~
{\as r^2\over R^3_{cr}}~\sim~\as^{1/2}\La^2r,\label{sm}
\ee
and we reproduce the linear correction to the potential at short distances.

It is worth emphasizing, however, that $R_{cr}\sim r^{1/3}$
and tends to zero with $r\rightarrow 0$. Thus, the dynamics described above implies that the vacuum is fine grained. Which is far from being obvious. Alternatively, we may say that if the linear correction to the potential
at short distances is confirmed this would have highly nontrivial implications for QCD.  

{\bf Tachyonic gluon mass.}

If we decide to look for a dynamical framework for a new short-distance physics in QCD, then monopoles seem to be a natural candidate (the topic of monopoles was extensively covered at this Workshop and for reviews see, e.g. \cite{monopole}). The monopole
s appear in the abelian projection and are usually discussed in connection with the confinement mechanism, or physics at large distances of
order $\La^{-1}$ (see, however, \cite{polikarpov}). Here, we are interested in the short-distance aspect
of the problem. Moreover, we would argue that if the monopoles are indeed an important field configuration in QCD, then a new short-distance physics is
introduced.  
Indeed, in theories with elementary Higgs fields, the singular behaviour of gauge fields at short distances is smeared. However, in QCD at short distances no composite field can be relevant and the singular behaviour 
of the gauge fields can be regularized only at the lattice size,
at least at first sight so. These are very interesting questions to our mind but they were not discussed, to our knowledge, in original papers and we will, therefore, confine ourselves to a few remarks sketching a possible new kind of phenomenology with a
n effective tachyonic gluon mass, $-\la_g^2<0$. This time the effect  
is coming from short distances and the $\la_g^2$ is not to be confused with
$\la^2$ considered in Section 2 in connection with the IR renormalon (or non-analytical in $\la^2$ terms). However,
there is a common feature of $\la^2$ and $\la_g^2$: in full nonabelian theory both can be consistently
tried only in one-loop calculations.

The abelian models with condensation of monopoles can be viewed as (dual) Higgs models, with two masses
$m_H^2$ and $\la^2$, of spin 0 and spin 1 particles, 
respectively. The quark static potential is then given by (see, e.g., \cite{suganuma}):
\be
V(r)~=~-{Q^2\over 4\pi}{e^{-\la r}\over r}+
{Q^2\la^2\over 8\pi}ln\left({\la^2+m_H^2\over\la^2}\right)\cdot r
\label{tension}
\ee
where $Q$ is the abelian charge. Moreover, the first term represents Yukawa-type
potential due to a (massive) gluon exchange while the second term represents
the confining potential $kr$, where $k$ is the string tension. 
As is mentioned above, the potential (\ref{tension}) is considered usually
at large $r$. For us, the central question is how far we can extend it to the region of small $r$. The answer is easy to obtain in the  the London limit,
$m_H^2\gg \la^2$. Then the linear piece in (\ref{tension}) 
can be trusted also at $\la^{-1}<r<m_H^{-1}$. This can be seen from 
more detailed explicit
expressions for the potential  \cite{suganuma} or one may simply observe that in the limit $m_H\rightarrow \infty$ the core of the string connecting the quarks becomes infinitely thin. Let us mention that there is numerical evidence 
\cite{suzuki} that the realistic QCD is close to the London limi but this is far from being a settled question (for further discussion see \cite{suzuki,monopole}). We simply assume $m_H\approx\Lambda_{UV}$ since it is the easiest way to implement zero siz
e of monopoles (see discussion at the beginning of the subsection), although other possibilities may also exist.

Then for the correction to the potential at short distances we get:
\be
\lim_{r\rightarrow~0}{\de V(r)}~=~-\la^2r\left({Q^2\over 8\pi}-
{Q^2\over 16\pi} ln\left({m_H^2\over\la^2}\right)\right)~\equiv~-{(-\la_g^2)
\over 2}r.\ee
Here $\la_g^2$ is positive and the expansion corresponds to an effective tachyonic
gluon mass $-\la_g^2$. Numerically, $\la_g^2\approx 3k/2\as$. Note that our simple-minded estimate (\ref{sm}) also corresponds to a 
positive $\de V(r)$ and, therefore, could be interpreted as due to a tachyonic gluon mass.

We could have tried to scrutinize the derivation but would prefer to jump directly at the phenomelogical manifestations of the tachyonic gluon mass. In fact the phenomenology turns not trivial at all and, to our knowledge, does not echo any other approach
.  
We illustrate this statement by example of the QCD sum rules in the $\rho$-meson and pion channels. As is well known \cite{nsvz1} these channels exhibit striking differences and, essentially, the pion channel cannot be understood in the standard terms of 
the IR sensitive power corrections.  

If one accounts for a non-vanishing gluon mass $\la$ then in one-loop approximation
the absorptive parts of the current correlators in the vector and pseudoscalar channels for massless quarks look as \footnote{I am deeply indebted to K.G. Chetyrkin for providing me with these equations.}:
\bea
Im\Pi_{\rho}(s)~=~1+{\as\over \pi}\\ \nonumber
Im\Pi_{\pi}(s)~=~1+{17\as\over 3\pi}+{2\as\over \pi} ln{\mu^2\over s}-{4\as\over \pi}{\la^2\over s}.\eea
Here $\mu$ is the normalization point and the term proportional to
$ln(\mu^2/s)$ is due to a non-vanishing anomalous dimension of the 
pseudoscalar current. Moreover, the currents are normalized in such a way that
for bare quarks ($\as=0$) the imaginary part is equal to unit. Note also
that the gluon condensate would manifest itself as a term proportional to
$\delta^{'}(s)$ while in the equations above $s\neq 0$. 

We are interested now in the effect of the gluon mass which affects only
the pseudoscalar channel and in accordance with our discussion above we
substitute a tachyonic mass instead of $\la^2$ so that a
rough estimate runs as follows:
\be
-{4\as\la^2\over\pi s}~\rightarrow~\approx +{2k\over s}~\approx~+{0.4 GeV^2\over s}.
\ee
Thus, the effect reaches $\sim$20\% at $s\sim 2GeV^2$ which fits very well,
both in the magnitude and sign, what
is needed  to bring the sum rules in agreement with the data, the agreement which is out of reach \cite{nsvz1} of the conventional phenomenology 
\footnote{There exist successful fits within the
model of the instanton liquid \cite{shuryak} (see also \cite{nsvz1}). It is not necessarily conflicting models since monopoles may be interconnected with the instantons (see \cite{monopole} and references therein).}.  
It is amusing that the correction $\sim\la^2$ does not appear in the vector
channel where there exist quite strict bounds  on the $1/s$
terms \cite{narison}.
Preliminary analysis shows
that the $\la_g^2$ could resolve the problems \cite{nsvz1} with sum rules  in the gluonic channels as well. It is worth mentioning that the preliminary evidence for the $1/Q^2$ correction on the lattice \cite{pino} could also
be fitted with a tachyonic mass for the gluon. 

{\bf Summary to part III.}

In this part we have considered dynamical models for nonperturbative effects in QCD, both in infrared and ultraviolet, motivated by multi-renormalon chains. In case of the IR renormalons we argued in favor of convergence of the renormalons with the old fa
shioned tube model. In case
of the UV physics, we speculated about applicability of the string model with
infinitely thin core. Phenomenologically, the model amounts to introduction
of a tachyonic effective mass for gluons and there are first indications that the model might resolve old standing problems with QCD sum rules.

{\bf ACKNOWLEDGMENTS.}

I am thankful to the Organizing Committee
of the YKIS97, and especially to Prof.
T. Suzuki, for the opportunity to participate in this
meeting. The review is based to a great extent on the papers written in
collaboration with Prof. R. Akhoury and I acknowledge gratefully numerous thorough
discussions with him of the topics covered in these lectures. I am also
grateful to Prof. M.I. Polikarpov for enlightening discussions.


\begin{thebibliography}{99} 
\bi{azimov}
G. 't Hooft, {\it in} "The whys of subnuclear physics",
Erice 1977, Ed. Zichichi, Plenum (1979), p. 94.
\bibitem{mueller}A.H. Mueller, {\it Nucl. Phys.} {\bf B250} (1985) 327 .
\bibitem{vz}
V.I. Zakharov, {\it Nucl. Phys.} {\bf B385} (1992) 452 .
\bibitem{sterman}H. Contopanagos and G. Sterman, {\it Nucl.
Phys.}  {\bf B419}, 77 (1994); G. Korchemsky and G. Sterman, {\it Nucl.Phys.} {\bf B437}, 415
(1995).
\bibitem{webber}B.R. Webber, {\it Phys. Lett.} {\bf B339} (1994) 148 
\bibitem{az}
R. Akhoury and V.I. Zakharov,
{\it Nucl.Phys.Proc.Suppl.} {\bf 54A} (1997) 217 (hep-ph/9610492); 
M. Beneke, hep-ph/9706457;
B. Webber, hep-ph/9712236. 
\bibitem{vainshtein}
A.I. Vainshtein and V.I. Zakharov, 
{\it Phys. Rev. Lett.} {\bf 73} (1994) 1207; {\it Phys. Rev.}
{\bf D54} (1996) 4039. 
\bibitem{dokshitzer}
Yu. L. Dokshitser and B.R. Webber, {\it Phys. Lett}
{\bf B352} (1995) 451;\\
Yu. L. Dokshitzer, G. Marcehsini and B. Webber,
{\it Nucl.Phys.} {\bf B469} (1996) 93\\
Yu.L. Dokshitser, B.R. Webber,  
{\it Phys.Lett.} {\bf B404} (1997) 321. 
\bibitem{az3} 
R. Akhoury and V.I. Zakharov, {\it Phys. Lett.}
{\bf B357}  (1995) 646; {\it Nucl. Phys. B}
{\bf B465} (1996) 295. 
\bibitem{sterman2}
G. Korchemsky and G. Sterman, hep-ph/9505391;        
G.P. Korchemsky, G. Oderda, G. Sterman
hep-ph/9711277 
\bi{bbz}
M. Beneke, V. Braun, and V.I. Zakharov, {\it Phys. Rev. Lett.}
{\bf 73} (1994) 3058.
\bi{stein}
E. Stein, M. Meyer-Hermann, L. Mankiewisz, A. Sch\"{a}fer,
{\it Phys. Lett.} {\bf B376} (1996) 177.
\bi{az2}
R. Akhoury and V.I. Zakharov, hep-ph/9701378.
\bi{bb}
M. Beneke and  V.M. Braun, {\it Nucl. Phys.} {\bf B454} (1995) 253.
\bibitem{az4} 
R. Akhoury and V.I. Zakharov, {\it Phys. Rev. Lett.}
{\bf 76} (1996) 2238;\\.R. Akhoury, L. Stodolsky, and V.I. Zakharov,
hep-ph/9609368. 
\bi{asz}
R. Akhoury, M.G. Sotiropoulos, and V.I. Zakharov,
{\it  Phys.Rev.} {\bf D56} (1997) 377. 
\bi{dd}
V. Del Duca, {\it Nucl. Phys.} {\bf B345} (1990) 369.
\bi{faleev}
S.V. Faleev and P.G. Silvestrov, hep-ph/9610344.
\bi{az5}
R. Akhoury and V.I. Zakharov, hep-ph/9705318.
\bi{az6}
R. Akhoury and V.I. Zakharov,  hep-ph/9710487. 
\bi{also}
N.V. Krasnikov and A.A. Pivovarov, 
{\it Mod. Phys. Lett.} {\bf A11} (1996) 835;\\ 
Yu.L. Dokshitser and N.G. Uraltsev, {\it Phys. Lett.}
{\bf B380} (1996) 141. \\ S. Peris and E. de Rafael,
{\it Phys. Lett.} {\bf B387} (1996) 603.
\bi{agl}
U. Aglietti and Z. Ligeti, {\it Phys. Lett.} {\bf B364} (1995) 75.
\bi{peter}
M. Peter, {\it Phys. Rev. Lett.} {\bf 78} (1997) 602.
\bi{twoloop}
Yu.L. Dokshitser, A. Lucenti, G. Marchesini, and G.P. Salam,
{\it Nucl. Phys.} {\bf B511} (1998) 396. 
\bi{pino}
G. Burgio, F. Di Renzo, G. Marchesini, and E. Onofri, hep-ph/9706209, 
hep-lat/9709105.  
\bibitem{bz}    
M. Beneke and V.I. Zakharov,  {\it Phys. Rev. Lett.}
{\bf 69} (1992) 2472. 
\bi{parisi}
G. Parisi, {\it Phys. Lett.} {\bf B76} (1978) 65.
\bibitem{kivel}
 M. Beneke, V.M. Braun, N. Kivel,
{\it Phys. Lett} {\bf B404} (1997) 315.   
\bi{peris}
S. Peris and  E. de Rafael, {\it  Nucl. Phys.}
{\bf B500} (1997).
\bi{yamawaki}
K. Yamawaki and V.I. Zakharov,  hep-ph/9406373. 
\bi{svz}
M.A. Shifman, A.I. Vainshtein, and V.I. Zakharov,
{\it Nucl. Phys.} {\bf B147} (1979) 385, 448. 
\bi{shifman} M. Shifman, hep-ph/9802214.  
\bi{grunberg}
G. Grunberg, hep-ph/9705290, hep-ph/9705460, hep-ph/9711481.    
\bi{aswell}
P.J. Redmond, {\it Phys. Rev.} {\bf 112} (1958) 1404;\\
D.V. Shirkov, hep-ph/9708480; hep-th/9709156; I.L. Solovtsov and D.V. Shirkov 
{\it Phys. Rev. Lett.} {\bf 79} (1997) 1209;\\ A.I. Alekseev and B.A. Arbuzov,
hep-ph/9704228;\\
E. Gardi and M.Karliner, hep-ph/9802218. 
\bibitem{webber3}
B.R.Webber, {\it Proc. Summer School on Hadronic Aspects
of Collider Physics, Zuoz, Switzerland}, ed. M.P.Locher (PSI, Villigen, 1994).
\bibitem{barreiro}
F. Barreiro, {\it Fortshr. Phys.} {\bf 34}, 503 (1986).
\bi{nason}
P. Nason and M.H. Seymour, \NP {\bf B454} (1995) 291.
\bi{delphi}
P. Abreu et al. (DELPHI Collaboration ), {\it Z. Phys.}
{\bf C73} (1997) 229. 
\bi{nsvz}
 V.A. Novikov, M.A. Shifman, A.I. Vainshtein, and  V.I. Zakharov,
\NP {\bf B174} (1980) 378. 
\bi{bali}
G. Bali, K. Schilling, and A. Wachter, hep-lat/9506017.
\bi{monopole}
M.N. Chernodub and M.I. Polikarpov, hep-th/9710205; 
A. Di Giacomo, hep-lat/9802008;
G.S. Bali, Ch. Schlichter, and K. Schilling, hep-lat/9802005. 
\bi{polikarpov}
 B.L.G. Bakker, M.N. Chernodub, and M.I. Polikarpov,
{\it Phys. Rev. Lett.} {\bf 80} (1998) 30.
\bi{suganuma}
H. Suganuma, S. Sasaki, and H. Toki, {\it  Nucl. Phys.} {\bf B435} (1995) 207.
\bi{suzuki}
T. Suzuki et. al., Proceedings of the Workshop  " Confinememnt, Duality and Nonperturbative Aspects of QCD", Cambridge (UK), June 1997.
\bi{nsvz1}
V.A. Novikov, M.A. Shifman, A.I. Vainshtein, and V.I. Zakharov, \NP
{\bf B191} (1981) 301.
\bi{narison}
S. Narison, {\it Phys. Lett.} {\bf B345} (1995) 166. 
\bi{shuryak}
T. Sch\"afer and E.V. Shuryak, hep-ph/9610451.
\end{thebibliography}
\end{document}